\documentclass[pra,aps,twocolumn,superscriptaddress,a4paper,floatfix,longbibliography]{revtex4-2}
\usepackage[latin1]{inputenc}
\usepackage{graphicx}
\usepackage{amsmath,amssymb}
\usepackage{hyperref}
\usepackage{cleveref}
\usepackage{gensymb}
\usepackage{mathtools}

\newcommand \beq{\begin{eqnarray}}
\newcommand \eeq{\end{eqnarray}}

\newcommand \mk{\mathbf{k}}

\begin{document}

\title{Tunneling Hamiltonian analysis of DC Josephson currents in a weakly-interacting Bose-Einstein condensate}

\author{Shun Uchino}
\affiliation{Advanced Science Research Center, Japan Atomic Energy Agency, Tokai 319-1195, Japan}

%\date{\pdfdate}

\begin{abstract}

Atomtronics experiments with ultracold atomic gases allow us to explore quantum transport phenomena of a weakly-interacting Bose-Einstein condensate (BEC).
Here, we  focus on  two-terminal transport of such a BEC in the vicinity of zero temperature.
By using the tunnel Hamiltonian and Bogoliubov theory, we obtain a DC Josephson current expression in the BEC
and apply it to experimentally relevant situations such as  quantum point contact and planar junction. 
Due to the absence of  Andreev bound states but the presence of couplings associated with condensation elements, a current-phase relation  in the BEC
is found to be different from one in an $s$-wave superconductor. 
In addition, it turns out that the DC Josephson current in the BEC depends on the sign of tunneling elements, which allows to realize the so-called $\pi$ junction
by using techniques of artificial gauge fields.

\end{abstract}

%\pacs{
%}

\maketitle

\section{introduction}
Quantum transport between macroscopic reservoirs 
reveals colorful quantum phenomena.
When macroscopic reservoirs are separated by a two-dimensional barrier,
a variety of quantum tunneling phenomena that depends on quantum states of matter in reservoirs emerge~\cite{duke}.
A prototypical example is a superconducting tunneling junction where an insulating barrier is sandwiched by superconductors.
In such a junction also known as the Josephson junction, electrons can flow even in the absence of a voltage bias, 
which is also an important building block of superconducting qubits~\cite{kjaergaard2020}.
 When macroscopic reservoirs  are connected by  mesoscopic samples,
in addition,  quantum nature in transport is known to be enhanced~\cite{datta1997}.
If reservoirs are filled in superconductors, for instance, the mesoscopic supercurrent is essentially determined by properties of  Andreev bound states~\cite{nazarov2009}.
In order to reveal such quantum transport phenomena, 
typical condensed matter systems composed of semiconductors and superconductors have conventionally been examined.

When it comes to a purely fundamental physics point of view, one may take advantage of controllable artificial systems to 
deepen understanding of quantum transport.
Ultracold atomic gases are one of such systems~\cite{amico2021}, and now allow to realize typical quantum transport systems such as
the tunneling junction~\cite{PhysRevLett.95.010402,levy2007ac,neri} and mesoscopic systems~\cite{PhysRevA.93.063619,krinner2017}.
An interesting perspective in transport research with ultracold atomic gases is to explore regimes
hard to attain with solid-state materials.

A weakly-interacting Bose-Einstein condensate (BEC) realized in ultracold atomic gases is a distinctive example in that
one can examine a variety of transport properties of a BEC in a controlled way.
Since the realization of a weakly-interacting BEC in ultracold atomic gases, numerous theoretical studies on tunneling phenomena between BECs
have been done in connection with the celebrated Josephson effect~\cite{PhysRevLett.79.4950,PhysRevA.59.620,RevModPhys.73.307}, which now becomes one chapter of the typical textbooks~\cite{pethick,pitaevskii2016}.
In addition to the theoretical works, the corresponding experimental works on the Josephson physics
have also been done~\cite{PhysRevLett.95.010402,levy2007ac,PhysRevLett.105.204101,PhysRevLett.106.130401,PhysRevLett.106.025302,valtolina2015,PhysRevLett.113.045305,PhysRevA.93.063619,PhysRevLett.123.260402,PhysRevLett.124.045301,kwon2020,luick2020,PhysRevLett.126.055301}.
\begin{figure}[htbp]
\centering
\includegraphics[width=8.5cm]{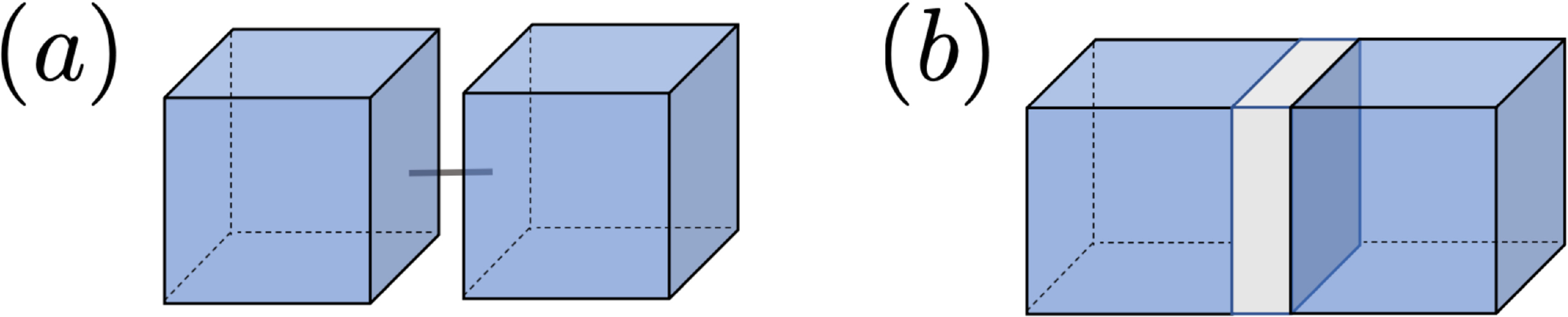}
\caption{\label{fig1} Two-terminal transport systems discussed in this work: (a) quantum point contact and (b) planar junction.}
\end{figure}

Josephson effects emerging in junction systems can be formulated in terms of the so-called
tunneling Hamiltonian~\cite{mahan2013}.
Then, a standard assumption is that tunneling amplitudes are so small that an analysis up to the linear response theory is justified.
However, it is not always trivial on whether or not experiments look at the linear response regime in reality, and 
it may also be important to estimate effects beyond the linear response theory~\cite{caroli1971,PhysRevLett.68.2512,PhysRevB.84.155414}.
In the typical mesoscopic setups such as a quantum point contact, moreover,
it is easy to reach the regime where higher-order tunneling effects are nonnegligible~\cite{zagoskin1998}.

The purpose of this paper is to obtain a DC Josephson current  of a weakly-interacting BEC that includes higher-order tunneling processes in the tunneling Hamiltonian.
To this end, we apply the Bogoliubov theory that explains a BEC near zero temperature to the tunneling Hamiltonian, and 
adopt the Keldysh formalism to take into account higher-order tunneling processes in an efficient manner.
As discussed in Refs.~\cite{PhysRevA.64.033610,PhysRevResearch.2.023284}, 
dominant contributions in low-energy transport between BECs  are tunneling processes associated with
condensation elements. Therefore, we focus on effects of such contributions, which is reasonable  near zero temperature, since the occupancy effect of Bogoliubov modes is negligible.
In consideration of cold-atom experiments, our formulation is specifically applied to quantum point contact and
planar junction (see also Fig.~\ref{fig1}).
We find that as increasing the tunnel coupling, the DC Josephson current in a BEC deviates from the conventional sinusoidal curve in 
a different way from one in an $s$-wave superconductor.
We note that our formulation is also applicable to other BEC junction systems such as spinor BECs in ultracold atomic gases~\cite{PhysRevLett.102.185301,RevModPhys.85.1191}  and magnon BECs in condensed matter~\cite{PhysRevB.90.144419}
if the corresponding quantum point contact and planar junction systems are realized.

This paper is organized as follows.
Section~II discusses the Bogoliubov theory and tunneling Hamiltonian, which give  a basis of formulation of
DC Josephson currents in a weakly-interacting BEC.
In Sec.~III, the Keldysh formalism that allows us to examine higher-order tunneling processes is 
applied to the tunneling Hamiltonian.
 Section~IV  obtains dominant contributions of DC Josephson currents in a BEC 
and analyzes quantum point contact and planar junction cases in a concrete manner. 
Section~V discusses related subjects such as internal Josephson effect, thermal current, trapped systems,
and possible experiments.
For a self-contained description, Green's function formulae used in quantum point contact and planar junction systems and
DC Josephson current formula in an $s$-wave superconductor are respectively given in Appendices A and B.

\section{Formulation of the problem}
We consider  two-terminal transport systems 
where two macroscopic reservoirs are weakly coupled via transport channels.
In our formulation, we assume that two reservoirs are in equilibrium in that
the thermodynamic variables such as temperature, chemical potential, and phase degree of freedom are fixed in time.
If the reservoirs are infinite, this assumption is reasonable.
In ultracold atomic gases where a finite system is concerned, however,
in the presence of a coupling between reservoirs,
the particle number in each reservoir can change in time.
Thus, the applicability of the above assumption may not  be trivial.

At the same time, in the typical mesoscopic setups with ultracold atomic gases such as quantum point contact and
tunneling junction, we encounter a situation that the thermalization time in each reservoir is much faster than the transport time
that is an equilibrated time of the total system including both reservoirs and channel.
In such a situation,
one may use the so-called quasi-steady approximation where two reservoirs are in equilibrium in each time of the evolution.
Since an analysis based on the quasi-steady approximation has a great success in cold atoms~\cite{krinner2017},
in what follows,  two-terminal transport properties in the presence of a fixed phase bias between reservoirs are discussed.

\subsection{Bogoliubov theory}
First, we focus on bulk properties of a weakly-interacting BEC.
The system of interest is that a number of non-relativistic identical bosons are trapped in each reservoir, where 
each atom experiences  an $s$-wave collision.
We can write down the corresponding reservoir Hamiltonian as follows:
\beq
H_{\alpha}=\sum_{\mk}[\epsilon_{\mk}-\mu]b^{\dagger}_{\mk,\alpha}b_{\mk,\alpha}+\frac{g}{2}\sum_{\mathbf{p,k,q}}b^{\dagger}_{\mathbf{p},\alpha}b^{\dagger}_{\mk,\alpha}b_{\mathbf{k-q},\alpha}b_{\mathbf{p+q},\alpha},
\label{eq:original}
\nonumber\\
\eeq
where $\alpha=L$ or $R$ is the reservoir index,
$\epsilon_{\mk}=k^2/2M$, and $\mu$ and $g$ are respectively the chemical potential and coupling constant, each of 
which takes a same value between reservoirs.
For the stability reason, the coupling $g$ is assumed to be positive, that is, repulsive interaction.
In this paper, we set $\hbar=k_B=1$ and  
the  volume in each reservoir is taken to be unity.

We now consider a situation that a temperature of the system is near absolute zero and therefore
bosons in each reservoir form a BEC.
Since each boson is weakly interacting, 
we can apply the so-called Bogoliubov prescription to the above Hamiltonian~\cite{fetter2012}.
Namely, we assume that the $\mathbf{k}=\mathbf{0}$ component gives the dominant contribution and
is replaced by $c$-number such that $b_{\mathbf{0},\alpha}=\sqrt{v_{\alpha}}e^{-i\phi_{\alpha}}$.
Here, $v_{\alpha}$ and $\phi_{\alpha}$ are physically the condensate fraction and phase degree of freedom in a BEC, respectively.
In the leading order analysis, the Hamiltonian is replaced by a $c$-number, which is  a function of $v_{\alpha}$.
Notice that up to the leading order, $\phi_{\alpha}$ dependence in the Hamiltonian is absent, since
the original Hamiltonian consists of the combination $b^{\dagger}b$ (U(1) symmetry). 
Then, $v_{\alpha}$ itself is determined from the stationary condition.
 By assuming the uniform solution, we obtain 
 \beq
 \mu=v_{\alpha}g.
 \eeq
 
 We next consider a fluctuation effect from the $c$-number analysis above.
 Namely,  we take into account $\mk\ne\mathbf{0}$ components that are indeed generated 
 in the presence of the interaction.
By using the stationary condition above, one can show that the term linear in $b_{\mathbf{k}\ne\mathbf{0},\alpha}$ vanishes.
Thus, the first nontrivial fluctuation effect comes from the term quadratic in $b_{\mathbf{k}\ne\mathbf{0},\alpha}$.
Up to this order, which is indeed the matter in the Bogoliubov theory,  we obtain the following effective Hamiltonian:
\beq
H_{\alpha}&&=\sum_{\mk\ne\mathbf{0}}[\epsilon_{\mk}+\mu]b^{\dagger}_{\mk,\alpha}b_{\mk,\alpha}\nonumber\\
&&\ \ +\frac{\mu}{2}\sum_{\mk\ne\mathbf{0}}[e^{-2i\phi_{\alpha}}b^{\dagger}_{\mk,\alpha}b^{\dagger}_{-\mk,\alpha}+e^{2i\phi_{\alpha}}b_{\mk,\alpha}b_{-\mk,\alpha}],
\label{eq:effective}
\eeq
where we neglect the constant term.
Since the above Hamiltonian contains $b_{\mk}b_{-\mk}$ and $b^{\dagger}_{\mk}b^{\dagger}_{-\mk}$ where
the explicit phase dependence appears,
one can introduce the following Bogoliubov transformation:
\beq
b_{\mk,\alpha}=e^{-i\phi_{\alpha}}u_k\beta_{\mk,\alpha}-
e^{-i\phi_{\alpha}}v_k\beta^{\dagger}_{-\mk,\alpha},
\eeq
with
\beq
u_k/v_k=\sqrt{\frac{\epsilon_{\mk}+\mu\pm E_{\mk}}{2E_{\mk}}},\\
E_{\mk}=\sqrt{\epsilon_{\mk}(\epsilon_{\mk}+2\mu)}.
\label{eq:energy}
\eeq
Then, 
the effective Hamiltonian is diagonalized as $H_{\alpha}=\sum_{\mk}E_{\mk}\beta^{\dagger}_{\mk,\alpha}\beta_{\mk,\alpha}$,
where we again neglect the constant term.
We note that as a result of the common chemical potential,
the Bogoliubov excitation energy $E_{\mk}$ does not depend on reservoirs.
Since the Bogoliubov transformation is canonical, 
it follows that $\beta_{\mk,\alpha}$ obeys the following commutation relation:
\beq
[\beta_{\mk,\alpha},\beta^{\dagger}_{\mk',\beta}]=\delta_{\alpha,\beta}\delta_{\mk,\mk'}.
\eeq
Below, Eqs.~\eqref{eq:effective}-\eqref{eq:energy} are used to obtain DC Josephson currents.

\subsection{Tunneling Hamiltonian}
We now introduce a coupling between reservoirs that induces transport.
We are particularly interested in a situation that such a coupling is described by the following tunneling term:
\beq
H_T&&=-\sum_{\mathbf{p},\mathbf{k}}(t_{\mathbf{p}\mk}b^{\dagger}_{\mathbf{p},R}b_{\mathbf{k},L}+t^*_{\mathbf{p}\mk}
b^{\dagger}_{\mathbf{k},L}b_{\mathbf{p},R}).
\eeq
Here $t_{\mathbf{p}\mk}$ is the tunneling amplitude that obeys $t^*_{\mathbf{p}\mk} =t_{-\mathbf{p}-\mk}$ in systems with time reversal symmetry.
Thus,  the Hamiltonian of the total system is described by sum of reservoirs' Hamiltonian and tunneling term
  $H=H_L+H_R+H_T$, which is the so-called tunneling Hamiltonian.
Josephson current analyses based on tunneling terms are successfully applied to discuss low-energy transport of tunnel junction~\cite{barone1982,PhysRevA.64.033610}, quantum point contact~\cite{PhysRevB.54.7366,PhysRevA.98.041601,PhysRevResearch.2.023284},
and quantum dot systems~\cite{martin2011}. 
 
 In the presence of $H_T$, the  particle number in each reservoir is not conserved and 
 a nonzero particle current is induced in the presence of a bias between reservoirs.
 By taking the convention that a current is positive when particles flow from left to right reservoirs,
the particle current operator in the tunneling Hamiltonian is  expressed as
\beq
I_N=-\dot{N}_L=i[N_{L},H]=i[N_{L},H_T],
\eeq
where $N_L=\sum_{\mk}b^{\dagger}_{\mk,L}b_{\mk,L}$ is the particle number operator in reservoir $L$,
and we use the Heisenberg equation.
By using the boson commutation relation, the particle current operator is reduced to
\beq
I_N&&=i\sum_{\mathbf{p},\mathbf{k}}( t_{\mathbf{p}\mk}b^{\dagger}_{\mathbf{p},R}b_{\mathbf{k},L}- t^*_{\mathbf{p}\mk}
b^{\dagger}_{\mathbf{k},L}b_{\mathbf{p},R}).
\label{eq:mass}
\eeq
We note that the above expression of $I_N$ is an operator relation and therefore is available regardless of 
quantum states in detail~\footnote{As in the case of superconductors, calculation of the current operator must be 
done by using the exact reservoir Hamiltonian~\eqref{eq:original}, rather than
the approximate one~\eqref{eq:effective} that may lead to some artifact on breaking the total number of particles. }.

We now take into account the fact that each reservoir is filled by a BEC, where there is the global U(1) phase degree of freedom 
$\phi_{L(R)}$.
Our interest is the current induced by a relative phase between BECs, $\Delta\phi=\phi_L-\phi_R$.
To this end, it is convenient to perform the  U(1) gauge transformation $b_{\mk, L(R)}\to e^{-i\phi_{L(R)}}b_{\mk,L(R)}$.
Then, $\phi_{L(R)}$ dependence present in Eq.~\eqref{eq:effective} vanishes, and instead
the tunneling term is transformed as follows:
\beq
H_T\to-\sum_{\mathbf{p},\mathbf{k}}( t_{\mathbf{p}\mk}e^{-i\Delta\phi}b^{\dagger}_{\mathbf{p},R}b_{\mathbf{k},L}+ t^*_{\mathbf{p}\mk}e^{i\Delta\phi}
b^{\dagger}_{\mathbf{k},L}b_{\mathbf{p},R}),\nonumber\\
\label{eq:tunnel}
\eeq
where the relative phase dependence is present.
It is then straightforward to confirm that due to this gauge transformation, $I_N$ also gains 
the corresponding phase factor $e^{\pm i\Delta \phi}$.

In what follows, we calculate the currents based on this transformed 
Hamiltonian~\footnote{One can obtain the same current formula without introducing the gauge transformation.
Then, one must take care of the fact that uncoupled Green's functions discussed in the Sec.~III and Appendix A
 depend explicitly on the phase degrees of freedom.}.

\section{Current expression with real time Green's functions}
In order to examine quantum transport properties of a BEC, we use the following Nambu representation of the field operator:
\beq
\hat{b}_{\mathbf{k},\alpha}(\tau)=
\begin{pmatrix}
b_{\mathbf{k},\alpha}(\tau) \\
b^{\dagger}_{\mathbf{-k},\alpha}(\tau)
\end{pmatrix}.
\eeq
By using this representation, the tunneling term is expressed as 
\beq
H_T&&=\sum_{\mathbf{p},\mk}\hat{b}^{\dagger}_{\mathbf{p},R}
\begin{pmatrix}
-t_{\mathbf{p}\mk}e^{-i\Delta\phi} & 0\\
0 &- t_{\mathbf{p}\mk}e^{i\Delta\phi}
\end{pmatrix}
\hat{b}_{\mk,L},
\eeq
where we use the property of time reversal symmetry.
In this work, we adopt the Keldysh formalism~\cite{rammer2007,kamenev2011} to calculate the DC Josephson current.
Under the Nambu representation, contour-ordered Green's function has the following
 $2\times2$ matrix structure:
\beq
\hat{G}_{\mk\mathbf{p},\alpha\beta}(\tau_{a},\tau'_b)=-i
 \langle T_c[\hat{b}_{\mathbf{k},\alpha}(\tau_a)\hat{b}^{\dagger}_{\mathbf{p},\beta}(\tau'_b)] \rangle,
\eeq
where $T_c$ is the contour ordering, and $a,b=\pm$ denotes upper ($+$) and lower ($-$) branches on the contour~\cite{rammer2007,kamenev2011}.
We note that the above average $\langle\cdots\rangle$ is taken under the total Hamiltonian $H$ including the tunneling term.
In particular, the component of $a=+$ and $b=-$ is lesser Green's function given by
\beq
&&\hat{G}^<_{\mk\mathbf{p},\alpha\beta}(\tau,\tau')\nonumber\\
&&=-i\begin{pmatrix}
\langle b^{\dagger}_{\mathbf{p},\beta}(\tau')b_{\mathbf{k},\alpha}(\tau)  \rangle  & \langle b_{\mathbf{-p},\beta}(\tau')
b_{\mathbf{k},\alpha}(\tau)  \rangle \\
\langle b^{\dagger}_{\mathbf{p},\beta}(\tau')b^{\dagger}_{\mathbf{-k},\alpha}(\tau)  \rangle  & \langle b_{\mathbf{-p},\beta}(\tau')b^{\dagger}_{\mathbf{-k},\alpha}(\tau)  \rangle 
\end{pmatrix},
\nonumber\\
\eeq
By means of lesser Green's function introduced above,
the average particle current can be expressed as
\beq
&&I_N(\tau)=\sum_{\mathbf{p},\mk}\text{Tr}\Big[\hat{\sigma}_z\begin{pmatrix}
-t_{\mathbf{p}\mk}e^{-i\Delta\phi} & 0\\
0 &-t_{\mathbf{p}\mk}e^{i\Delta\phi}
\end{pmatrix} \hat{G}^<_{\mk\mathbf{p},LR}(\tau,\tau)\Big],\nonumber\\
\label{eq:current-green}
\eeq
where $\hat{\sigma}_z=\begin{pmatrix}1 & 0\\ 0 &-1  \end{pmatrix}$ and the trace is taken over the Nambu space.
For later use, we also introduce retarded and advanced Green's functions
\beq
&&\hat{G}^R_{\mk\mathbf{p},\alpha\beta}(\tau,\tau')=-i\theta(\tau-\tau')\langle [\hat{b}_{\mk,\alpha}(\tau),\hat{b}^{\dagger}_{\mathbf{p},\beta}(\tau') ]\rangle,\\
&&\hat{G}^A_{\mk\mathbf{p},\alpha\beta}(\tau,\tau')=i\theta(\tau'-\tau)\langle [\hat{b}_{\mk,\alpha}(\tau),\hat{b}^{\dagger}_{\mathbf{p},\beta}(\tau') ]\rangle.
\eeq

In the absence of $H_T$, each reservoir is decoupled and 
Green's functions can be determined by using the Bogoliubov theory.
If we denote the corresponding uncoupled Green's functions $\hat{g}$,
whose derivation is given in Appendix A,
the Dyson equation allows us to relate $\hat{G}$ with $\hat{g}$
by treating $H_T$ as perturbation.

Since $H_T$ is the single-particle potential term, 
the Dyson equation for $G^{R(A)}$ has the following simple structure~\cite{rammer2007}: 
\beq
\hat{G}^{R(A)}=\hat{g}^{R(A)}+\hat{g}^{R(A)}\circ \hat{V}\circ\hat{G}^{R(A)}.
\eeq
Here,
$\hat{V}$ represents $\hat{t}_{\mathbf{p}\mk}=\delta(\tau-\tau')
\begin{pmatrix}
-t_{\mathbf{p}\mk}e^{-i\Delta\phi} & 0\\
0 & -t_{\mathbf{p}\mk}e^{i\Delta\phi}
\end{pmatrix}
$ or $\hat{t}^{\dagger}_{\mk\mathbf{p}}$,
 and $\circ$ represents summation over the internal momentum variable and
  integration over the internal time variable from minus infinity to plus infinity.
On the other hand,  by using the so-called Langreth rules~\cite{rammer2007},
the Dyson equation for $\hat{G^<}$ is obtained as
\beq
\hat{G}^<=(\hat{1}+\hat{G}^R\circ\hat{V})\circ\hat{g}^<\circ(\hat{1}+\hat{V}\circ\hat{G}^A),
\eeq
where $\hat{1}=\delta_{\mathbf{p}\mk}\delta(\tau-\tau')\begin{pmatrix}1 &0 \\ 0 & 1  \end{pmatrix}$.

In order to obtain a convenient expression of the current, we now make the reservoir indices explicit.
Then, the Dyson equations for $\hat{G}^{R,A}$ are expressed as
\beq
&&\hat{G}^{R(A)}_{LR}=\hat{g}^{R(A)}_{LL}\circ\hat{t}^{\dagger} \circ\hat{G}^{R(A)}_{RR},\label{eq1} \\
&&\hat{G}^{R(A)}_{RL}=\hat{g}^{R(A)}_{RR}\circ\hat{t}\circ\hat{G}^{R(A)}_{LL}, \label{eq2}\\
&&\hat{G}^{R(A)}_{LL}=\hat{g}^{R(A)}_{LL}+\hat{g}^{R(A)}_{LL}\circ\hat{t}^{\dagger} \circ  \hat{g}^{R(A)}_{RR}\circ\hat{t}
\circ\hat{G}^{R(A)}_{LL},\label{eq3}\\
&&\hat{G}^{R(A)}_{RR}=\hat{g}^{R(A)}_{RR}+
\hat{g}^{R(A)}_{RR}\circ\hat{t}\circ  \hat{g}^{R(A)}_{LL}\circ
\hat{t}^{\dagger}\circ\hat{G}^{R(A)}_{RR},\label{eq4}
\eeq
where $\hat{t}$ is the short hand notation of $\hat{t}_{\mathbf{pk}}$.
In addition, the Dyson equation for $\hat{G}^<$ is rewritten as
\beq
\hat{G}^<_{LR}=\hat{G}^{<,{\text{odd}}}_{LR}+\hat{G}^{<,\text{even}}_{LR},
\eeq
where
\begin{widetext}
\beq
\hat{G}^{<,\text{odd}}_{LR}
&&=(\hat{1}+\hat{G}^R_{LR}\circ\hat{t})\circ \hat{g}^<_{LR}  \circ(\hat{1}
 +\hat{t}\circ\hat{G}^A_{LR}) +\hat{G}^R_{LL}\circ\hat{t}^{\dagger}\circ\hat{g}^<_{RL} \circ\hat{t}^{\dagger}\circ\hat{G}^A_{RR},
\label{eq5}
\eeq
\beq
\hat{G}^{<,\text{even}}_{LR}
&&=(\hat{1}+\hat{G}^R_{LR}\circ\hat{t}) \circ \hat{g}^<_{LL}\circ\hat{t}^{\dagger}\circ\hat{G}^A_{RR}
+\hat{G}^R_{LL}\circ\hat{t}^{\dagger}\circ\hat{g}^<_{RR}\circ(\hat{1}+\hat{t}\circ\hat{G}^A_{LR}).\label{eq6}
\eeq
In the above expressions,  odd and even in the superscript respectively represent current components of odd-order and even-order terms in tunneling amplitudes, 
which can be confirmed by looking at the Dyson equation for $\hat{G}^{R(A)}$ and Eq.~\eqref{eq:current-green} that contains
an additional tunneling amplitude dependence. We note that $\hat{G}^{<,\text{odd}}$ and $\hat{G}^{<,\text{even}}$ stem from 
$\hat{g}^<_{LR(RL)}$ and $\hat{g}^<_{LL(RR)}$ contributions, respectively.
What is peculiar in the BEC system is the presence of $\hat{g}^<_{LR(RL)}$ that arises from the $c$-number component of the field operator,
and therefore $\hat{g}^{R(A)}_{LR(RL)}=0$, since $\hat{g}^{R(A)}$ consists of the commutator of the field operators (see also Appendix A).
The presence of  $\hat{G}^{<,\text{odd}}$ is allowed by virtue of the tunnel coupling between condensation elements and one between condensation and non-condensation elements, and therefore  $\hat{G}^{<,\text{odd}}$ is absent in fermionic superconductors (see also Appendix B).
This difference between  fermionic  and bosonic systems can also be understood as 
the fact that  while in fermionic superconductors a phase degree of freedom responsible for Josephson currents
emerges from condensation of Cooper pairs whose transport is associated with even-order processes in the tunneling amplitude,
in the bosonic condensate discussed in this paper, odd-order terms in the tunneling amplitude are allowed 
owing to the property $\langle b\rangle\ne0$~\footnote{Odd-order terms in $t$ may also be absent in Bose liquids 
such that particle condensation is absent but pair condensation occurs~\cite{nozieres}.}.

For convenience, we also introduce renormalized tunneling matrices defined as
\beq
\hat{T}_{RL}^{R(A)}=\hat{t}+\hat{t}\circ\hat{g}^{R(A)}_{LL} \circ
 \hat{t}^{\dagger}\circ\hat{g}^{R(A)}_{RR}\circ \hat{T}_{RL}^{R(A)},\label{eq:tunnel-rl-t}\\
 \hat{T}_{LR}^{R(A)}=\hat{t}^{\dagger}+\hat{t}^{\dagger} \circ\hat{g}^{R(A)}_{RR} \circ
 \hat{t}\circ\hat{g}^{R(A)}_{LL} \circ \hat{T}_{LR}^{R(A)}\label{eq:tunnel-lr-t}.
\eeq
By using above, it is straightforward to show the following relations:
\beq
&&\hat{G}^{R(A)}_{LL}\circ\hat{t}^{\dagger}=\hat{g}^{R(A)}_{LL}\circ\hat{T}^{R(A)}_{LR},\\
&&\hat{t}^{\dagger}\circ\hat{G}^{R(A)}_{RR}=\hat{T}^{R(A)}_{LR}\circ\hat{g}^{R(A)}_{RR},\\
&&\hat{t}+\hat{t}\circ\hat{G}^{R(A)}_{LR}\circ\hat{t}=\hat{T}^{R(A)}_{RL}.
\eeq
Thus, we find
\beq
\text{Tr}\Big[\hat{\sigma}_{z}\hat{t}\circ\hat{G}^{<,\text{odd}}_{LR}\Big]&&=
\text{Tr}\Big[\hat{\sigma}_z\hat{T}^R_{RL}\circ\hat{g}^<_{LR}\circ(\hat{1}+\hat{t}\circ\hat{G}^A_{LR} )\Big] 
+\text{Tr}\Big[\hat{\sigma}_z\hat{t}\circ\hat{g}^R_{LL}\circ\hat{T}^R_{LR}\circ\hat{g}^<_{RL}\circ
\hat{T}^A_{LR}\circ\hat{g}^A_{RR}\Big], 
\label{eq:green1}
\eeq
\beq
\text{Tr}\Big[\hat{\sigma}_z\hat{t}\circ\hat{G}^{<,\text{even}}_{LR}\Big]
&&=\text{Tr}\Big[\hat{\sigma}_z\hat{T}^R_{RL}\circ\hat{g}^<_{LL}\circ\hat{T}^A_{LR}\circ\hat{g}^A_{RR}\Big]
+\text{Tr}\Big[\hat{\sigma}_z\hat{g}^R_{LL}\circ\hat{T}^R_{LR}\circ\hat{g}^<_{RR}\circ\hat{T}^A_{RL}\Big].
\label{eq:green2}
\eeq
The equations above give a foundation of DC Josephson currents in a weakly-interacting BEC.

\end{widetext}
\section{results}
 By using Eqs.~\eqref{eq:green1} and \eqref{eq:green2}, we show typical behaviors of DC Josephson currents in the vicinity of zero temperature, where
occupation effects of the Bogoliubov mode are negligible.
As theoretically discussed in Refs.~\cite{PhysRevA.64.033610,PhysRevA.100.063601,PhysRevResearch.2.023284,PhysRevResearch.2.023340} and experimentally 
verified in Refs.~\cite{kwon2020,PhysRevLett.126.055301}, the dominant contribution of a mesoscopic BEC in such a regime originates from
couplings associated with the condensation elements.
We point out that in the Keldysh formalism presented in this paper such a dominant contribution is extracted from the replacement $\hat{g}^<_{\mathbf{k},\alpha\beta}\to\hat{g}^<_{\mathbf{0},\alpha\beta}$.
In addition, since there is no chemical potential bias between reservoirs, an AC Josephson current is absent and
the current does not depend on time.
Thus, convenient expressions can be obtained by using the Fourier transformation.
In frequency space,  the DC Josephson current is expressed as
\beq
I_N\to\int_{-\infty}^{\infty} \frac{d\omega}{2\pi}[I_N^{\text{odd}}(\omega)+I_N^{\text{even}}(\omega) ]
\label{eq:main1}
\eeq
where $\to$ represents the prescription that the dominant contribution is picked up.
The odd-order and even-order terms in tunneling amplitudes are respectively expressed as
\begin{widetext}
\beq
I^{\text{odd}}_N(\omega)&&=\sum_{\mathbf{p},\mk}\text{Tr}\Big[\hat{\sigma}_z\hat{T}^R_{\mathbf{p}\mathbf{0},RL}(\omega)\hat{g}^<_{\mathbf{0},LR}(\omega)(\hat{1}\delta_{\mathbf{0}\mathbf{p}}+
\hat{t}_{\mathbf{0}\mk}\hat{G}^A_{\mk\mathbf{p},LR}(\omega)  )\Big]\nonumber\\
&&\ \ +\sum_{\mathbf{p},\mk}\text{Tr}\Big[\hat{\sigma}_z\hat{t}_{\mathbf{p}\mk}\hat{g}^R_{\mk,LL}(\omega)\hat{T}^R_{\mk\mathbf{0},LR}(\omega)\hat{g}^<_{\mathbf{0},RL}(\omega)
\hat{T}^A_{\mathbf{0}\mathbf{p},LR}(\omega)\hat{g}^A_{\mathbf{p},RR}(\omega)\Big],
\label{eq:main2}
\eeq
\beq
I^{\text{even}}_N(\omega)=\sum_{\mathbf{p}}\text{Tr}\Big[\hat{\sigma}_z\hat{T}^R_{\mathbf{p}\mathbf{0},RL}(\omega)\hat{g}^<_{\mathbf{0},LL}(\omega)
\hat{T}^A_{\mathbf{0}\mathbf{p},LR}(\omega)\hat{g}^A_{\mathbf{p},RR}(\omega)\Big]
+\sum_{\mk}\text{Tr}\Big[\hat{\sigma}_z\hat{g}^R_{\mk,LL}(\omega)\hat{T}^R_{\mk\mathbf{0},LR}(\omega)\hat{g}^<_{\mathbf{0},RR}(\omega)\hat{T}^A_{\mathbf{0}\mk,RL}\Big],
\label{eq:main3}
\eeq
\end{widetext}
where $\hat{1}=\begin{pmatrix}1 & 0\\ 0 &1 \end{pmatrix}$ and
$\hat{t}_{\mathbf{p}\mk}=\begin{pmatrix}-t_{\mathbf{p}\mk}e^{-i\Delta\phi} & 0 \\ 0 & -t_{\mathbf{p}\mk}e^{i\Delta\phi} \end{pmatrix}$ in the frequency-space representation.
Equations~\eqref{eq:main1}-\eqref{eq:main3} are the main results of this paper, which are applicable to generic tunneling amplitudes
within the perturbative meaningful regime.
In what follows, DC Josephson currents in point-contact and planar junction systems, which are relevant to cold-atom experiments, are separately discussed.

\subsection{Quantum point contact case}
Here we discuss the case of a quantum point contact where a particle transfer between reservoirs occurs through a one-dimensional wire.
The tunneling Hamiltonian description of a quantum point contact is known to be reasonable when short one-dimensional wire with channel transmittance whose 
energy dependence is negligible is concerned~\cite{PhysRevB.54.7366,PhysRevB.84.155414,husmann2015,PhysRevLett.118.105303,PhysRevA.98.041601,PhysRevA.100.043604,PhysRevResearch.2.023152,PhysRevResearch.2.023284,PhysRevResearch.2.023340}.
In particular, the tunneling term in a quantum point contact is modeled in such a way that the tunneling occurs at a single point in each reservoir, e.g.  at origin~\cite{husmann2015}.
In the momentum space representation, the corresponding tunneling term is expressed as
\beq
-t\sum_{\mathbf{p},\mk}(e^{-i\Delta\phi}b^{\dagger}_{\mathbf{p},R}b_{\mk,L}+^{i\Delta\phi}b^{\dagger}_{\mk,L}b_{\mathbf{p},R}),
\eeq
with the tunneling amplitude $t$.
Thus, compared with the general expression~\eqref{eq:tunnel},
 a momentum-independent tunneling amplitude  is now concerned in transport of a quantum point contact.

Since the momentum dependence in the tunneling amplitude is absent, 
the renormalized tunneling matrices become momentum independent. 
By using the frequency-space representation, we find that Eqs~\eqref{eq:tunnel-rl-t} and~\eqref{eq:tunnel-lr-t} are expressed as
\beq
\hat{T}^{R(A)}_{RL}(\omega)=\hat{t}+\hat{t}\hat{g}^{R(A)}_{LL}(\omega)\hat{t}^{\dagger}\hat{g}^{R(A)}_{RR}(\omega)\hat{T}^{R(A)}_{RL}(\omega),\\
\hat{T}^{R(A)}_{LR}(\omega)=\hat{t}^{\dagger}+\hat{t}^{\dagger}\hat{g}^{R(A)}_{RR}(\omega)\hat{t}\hat{g}^{R(A)}_{LL}(\omega)\hat{T}^{R(A)}_{LR}(\omega),
\eeq
where $\hat{t}=-t\begin{pmatrix}e^{-i\Delta\phi} & 0\\0 &e^{i\Delta\phi} \end{pmatrix}$ and $\hat{g}^{R(A)}_{\alpha\alpha}(\omega)$ is uncoupled local retarded (advanced) Green's function
whose expression is given in Appendix A.
Since the above are $2\times 2$ matrix equations, they can be solved as
\begin{widetext}
\beq
\hat{T}^{R(A)}_{RL}(\omega)=\frac{-t}{D^{R(A)}(\omega,\Delta\phi) }\begin{pmatrix}
e^{-i\Delta\phi}-t^2\{e^{-i\Delta\phi}(g^{R(A)}_{2} )^2+ e^{i\Delta\phi}(f^{R(A)} )^2\} & t^2(e^{i\Delta\phi}g_1^{R(A)}f^{R(A)} +e^{-i\Delta\phi}g_2^{R(A)}f^{R(A)}   )  \\
t^2(e^{i\Delta\phi}g_1^{R(A)}f^{R(A)} +e^{-i\Delta\phi}g_2^{R(A)}f^{R(A)}   )  & e^{i\Delta\phi}-t^2\{e^{i\Delta\phi}(g^{R(A)}_{1} )^2+ e^{-i\Delta\phi}(f^{R(A)} )^2\}
\end{pmatrix}, \nonumber\\
\eeq
\beq
\hat{T}^{R(A)}_{LR}(\omega)=\frac{-t}{D^{R(A)}(\omega,-\Delta\phi)}\begin{pmatrix}
e^{i\Delta\phi}-t^2\{e^{i\Delta\phi}(g^{R(A)}_{2} )^2+ e^{-i\Delta\phi}(f^{R(A)} )^2\} & t^2(e^{-i\Delta\phi}g_1^{R(A)}f^{R(A)} +e^{i\Delta\phi}g_2^{R(A)}f^{R(A)}   )  \\
t^2(e^{-i\Delta\phi}g_1^{R(A)}f^{R(A)} +e^{i\Delta\phi}g_2^{R(A)}f^{R(A)}   )  & e^{-i\Delta\phi}-t^2\{e^{-i\Delta\phi}(g^{R(A)}_{1} )^2+ e^{i\Delta\phi}(f^{R(A)} )^2\}
\end{pmatrix}, \nonumber\\
\eeq
where 
\beq
\hat{D}^{R(A)}(\omega,\Delta\phi)=t^4\{g^{R(A)}_1g^{R(A)}_2-(f^{R(A)})^2 \}^2-t^2\{(g^{R(A)}_1 )^2+(g^{R(A)}_2 )^2+2(f^{R(A)})^2\cos(2\Delta\phi) \}+1,
\eeq
and $g^{R(A)}_{1}$, $g^{R(A)}_2$, $f^{R(A)}$ are respectively 11, 22, 12 matrix elements of $\hat{g}^{R(A)}$. 
In addition, by using lesser Green's function expression shown in Eq.~\eqref{eq:lesser}, we find that  the dominant contribution of the DC Josephson current in a quantum point contact is obtained as
\beq
 I_N&&=-\frac{\mu i}{g}\text{Tr}\Big[\hat{\sigma}_z\hat{T}^R_{RL}(0)
\begin{pmatrix}
1 & 1\\
1 & 1
\end{pmatrix}
\hat{T}^A_{RL}(0)\hat{t}^{-1}
+\hat{\sigma}_z\hat{t}\hat{g}^R_{LL}(0)\hat{T}^R_{LR}(0)
\begin{pmatrix}
1 & 1\\
1 & 1
\end{pmatrix}
\hat{T}^A_{LR}(0)\hat{g}^A_{RR}(0)\nonumber\\
&&\ \ \ +\hat{\sigma}_z\hat{T}^R_{RL}(0)
\begin{pmatrix}
1 & 1\\
1 & 1
\end{pmatrix}
\hat{T}^A_{LR}(0)\hat{g}^A_{RR}(0) 
+ \hat{\sigma}_z\hat{g}^R_{LL}(0)\hat{T}^R_{LR}(0)
\begin{pmatrix}
1 & 1\\
1 & 1
\end{pmatrix}
\hat{T}^A_{RL} (0)
  \Big]\nonumber\\
  &&=\frac{2\mu t[1-t^2(\Phi(0)-\Psi(0))^2 ]\sin\Delta\phi }{g[1-t^2(\Phi(0)^2-\Psi(0)^2) +2t\Psi(0)\cos\Delta\phi]^2},
  \label{eq:leading}
 \eeq
where $\hat{t}^{-1}=-\frac{1}{t}\begin{pmatrix} e^{i\Delta\phi} & 0\\ 0 & e^{-i\Delta\phi} \end{pmatrix}$, and
$\Phi(0)$ and $\Psi(0)$ are real parts of $g_1^R(0)$ and $f^R(0)$, respectively (see also Appendix A).
\end{widetext}

We now comment on basic features of the DC Josephson current formula above.
First, as is expected from generic properties of Josephson currents,
Eq.~\eqref{eq:leading} is an odd function in $\Delta\phi$, that is, $I_{N}(-\Delta\phi)=-I_N(\Delta\phi)$.
Second, in contrast to fermionic superconductor and superfluid systems, where
Josephson currents consist of even-order terms in $t$,
$I_N$  above contains odd-order terms in $t$.
The presence of odd-order terms in $t$ shows that the current depends on the sign of $t$.
However, we point out that  the effect of the sign of $t$ can be absorbed by the following shift of the relative phase: $\Delta\phi\to\Delta\phi+\pi$.
Thus, it turns out that the result of a negative $t$ can be inferred from one of a positive $t$.

To see some intuition of Eq.~\eqref{eq:leading}, we next perform the series expansion in  $t$.
The leading term in Eq.~\eqref{eq:leading} is linear in $t$ and is given by
\beq
\frac{2\mu t\sin\Delta\phi}{g}.
\label{eq:josephson}
\eeq
Up to this order, the current obeys the conventional sinusoidal current-phase relation. 
We note that the result above can also be
obtained with substitution of the macroscopic wave function for $b^{\dagger}_{\mk=\mathbf{0},L(R)}$ and 
$b_{\mk=\mathbf{0},L(R)}$ in Eq.~\eqref{eq:mass}.  
We next look at the sub-leading term in Eq.~\eqref{eq:leading}, which is quadratic in $t$.
From the series expansion of Eq.~\eqref{eq:leading}, it is obtained as
\beq
-\frac{4\mu t^2\Psi(0)\sin2\Delta\phi}{g}.
\eeq
Thus, the argument of the sine function in this term is not $\Delta\phi$ but $2\Delta\phi$.
We point out that the essentially same result can also be obtained by using the linear response theory, where
the tunnel current up to $t^2$ is concerned.
It may also be clear from the analysis above that higher-order terms in $t$ generate current components proportional to $\sin n\Delta\phi$ with $n\ge3$.

\begin{figure}[htbp]
\centering
\includegraphics[width=8cm]{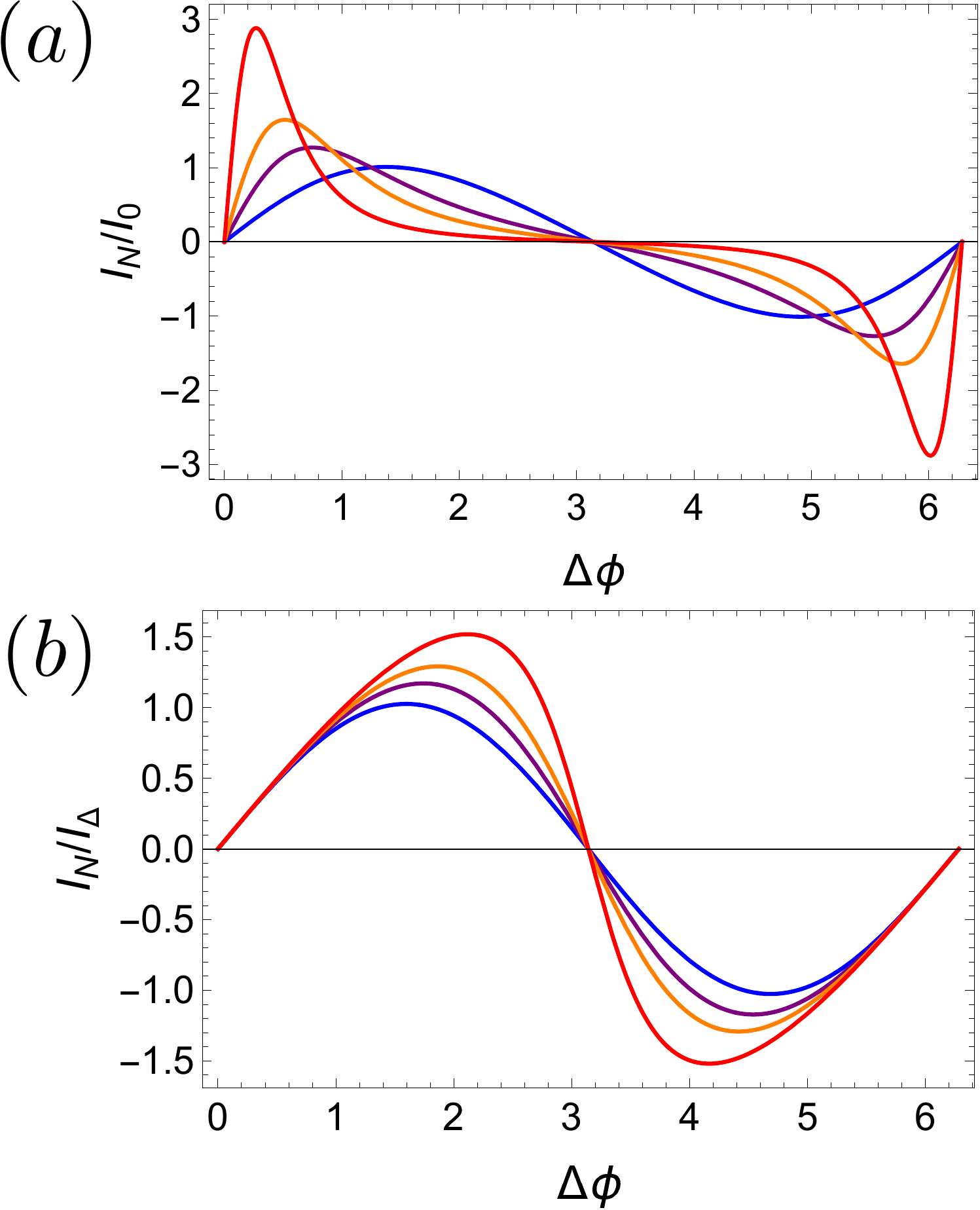}
\caption{\label{fig2} DC Josephson currents in quantum point contact systems. (a) The current-phase relation of a BEC
in units of  $I_0=2\mu t/g$ where $\Phi(0)=-\Psi(0)$ is assumed. The blue, purple, orange, and red curves are results at 
$|t\Psi(0)|=0.1$, $0.5$, $0.7$, and $0.9$, respectively.  As increasing the tunneling amplitude, 
the deviation from the sinusoidal curve becomes significant.
In (b), for comparison, we plot the current-phase relation of a conventional $s$-wave superconductor where the current is carried by
the Andreev bound state. There, the current is plotted in units of $I_{\Delta}={\cal T} \Delta/2$ with the superconducting gap $\Delta$ 
and the transmission parameter ${\cal T}$ defined in Appendix B. 
The blue, purple, orange, and red curves are respectively results at ${\cal T}=0.1$, $0.5$, $0.7$, and $0.9$.}
\end{figure}

We now directly look at  Eq.~\eqref{eq:leading}, which contains all order effects in $t$.
Figure~\ref{fig2}~(a) shows typical behaviors of the current-phase relation at different tunneling amplitudes.
Here the current is plotted in units of $I_0=2\mu t/g$ with a positive $t$, and the horizontal axis is restricted as $0\le\Delta\phi<2\pi$
due to the $2\pi$ periodicity.
In addition, we set $\Phi(0)=-\Psi(0)$, since $\Phi(0)$ contains the cutoff $\Lambda$ but we check that the value of $\Lambda$ is irrelevant to the qualitative discussion of the current-phase relation. Owing to this setting, the current-phase relation in a BEC can be characterized by a single dimensionless parameter $|t\Psi(0)|$.
We note that the condition $|t\Psi(0)|<1/2$ is necessary to obtain a perturbatively meaning result.
In terms of the physical model parameters, the current can also be expressed as
\beq
I_N=\frac{2\mu t[1-(\frac{t\mu^2}{c^3\pi})^2 ]\sin\Delta\phi }{g[1+\frac{t\mu^2}{c^3\pi}\cos\Delta\phi]^2},
\eeq
with the speed of sound $c=\sqrt{\frac{\mu}{M}}$.

In a small tunneling amplitude (blue curve), the current-phase relation obeys the conventional Josephson relation~Eq.~\eqref{eq:josephson}.
In larger tunneling amplitudes such as purple, orange, and red curves, on the other hand, 
we can see clear deviations from the conventional relation.  
There, the current takes large values in the vicinity of $\Delta\phi=0$ and $2\pi$, 
and is suppressed except for them.
Such a tendency is somewhat akin to one obtained by the Gross-Pitaevskii analysis with a barrier potential~\cite{PhysRevLett.99.040401,PhysRevA.79.033627,PhysRevA.81.033613,PhysRevResearch.2.033298}.
At the same time, we note that due to symmetry of the sign in $t$ mentioned before, the current-phase relation for a negative $t$ 
is shifted by $\pi$  and its amplitude takes large values in the vicinity of $\Delta\phi=\pi$.
In ultracold atomic gases, the current-phase relation for a negative $t$ that shows the $\pi$-junction behavior  may also 
be achieved by using techniques of synthetic gauge fields~\cite{aidelsburger2018}.

For comparison, we show the DC Josephson current in an $s$-wave superconductor in Fig.~\ref{fig2}~(b) (derivation with the tunneling Hamiltonian is given in Appendix B).
In the case of a superconductor, the DC Josephson current is carried by  tunneling of Cooper pairs via multiple Andreev reflections.
Due to the absence of a chemical potential bias, the loop of multiple Andreev reflections is closed and leads to the presence of Andreev bound states.
As in the case of a BEC, the Josephson current in a superconductor deviates from the conventional sinusoidal curve as increasing the tunneling amplitude.
This deviation from the conventional curve now originates from multiple Andreev reflections.
In contrast to the BEC case, as increasing the channel transmission parameter, the Josephson current becomes maximum near $\Delta\phi=\pi$, which does
not depend on the sign of the tunneling amplitude due to the absence of $\hat{G}^{<.\text{odd}}$ in a superconductor.

\subsection{Planar junction case}
We next consider the planar junction system where there is a translational invariance along the interface.
As pointed out in Ref.~\cite{PhysRevA.64.033610}, one can introduce three independent tunneling couplings in such a junction:
coupling between condensation elements, one between condensation and non-condensation elements, and one between non-condensation elements.
The corresponding tunneling amplitudes can respectively be expressed as~\cite{PhysRevA.64.033610}
\beq
&&t_{\mathbf{0}\mathbf{0}}=t_{cc}\delta_{\mathbf{0}\mathbf{0}},\\
&&t_{\mathbf{p}\mathbf{0}}=t_{nc}p\delta_{\mathbf{p}_{\parallel}\mathbf{0}},\\
&&t_{\mathbf{p}\mk}=t_{nn}pk\delta_{\mathbf{p}_{\parallel}\mk_{\parallel}},
\eeq
where $t_{cc}$, $t_{nc}$, and $t_{nn}$ are real constants, and $p$ and $k$ are momenta 
perpendicular to the interface~\footnote{In Ref.~\cite{PhysRevA.64.033610}, an additional momentum dependence thats lead to a saturation of tunneling amplitudes is introduced. 
In this paper, instead we introduce a sharp cutoff to regularize the results.}.
In $t_{\mathbf{00}}$, $\delta_{\mathbf{00}}$ represents the tunneling between zero momentum elements.
In addition, the factor $\delta_{\mathbf{p}_{\parallel}\mk_{\parallel}}$ in the coupling between non-condensation elements originates from the momentum conservation along the interface.

Our interest is the dominant contribution where $\mk=\mathbf{0}$ component of lesser Green's function is concerned.
Then, the momentum conservation along the interface leads to the fact that the three dimensional momentum summation appearing in the current expression is reduced to 
one dimensional one perpendicular to the interface.
This actually simplifies the problem and allows us to obtain the DC Josephson current expression of the planar junction in an analytic manner.

At the same time, the generic expression in the planar junction is still complex due to the presence of the three parameters $t_{cc}$, $t_{nc}$, and $t_{nn}$.
From a practical point of view, furthermore, an order-by-order analysis in tunnel couplings may be sufficient to discuss transport properties of the junction system.

Therefore, we turn to obtain the DC Josephson current of the planar junction, which contains first few series in tunneling amplitudes.
To be specific, we consider the current expression up to fourth order in tunneling amplitudes.
For this purpose, we can use the following approximations:
\beq
&&\hat{T}^{R(A)}_{\mathbf{p}\mk,RL}(\omega)\approx\hat{t}_{\mathbf{p}\mk}+\sum_{\mk_1,\mathbf{p}_1}\hat{t}_{\mathbf{p}\mk_1}\hat{g}^{R(A)}_{\mk_1,LL}(\omega)\hat{t}^{\dagger}_{\mk_1\mathbf{p}_1}\hat{g}^{R(A)}_{\mathbf{p}_1,RR}(\omega)\hat{t}_{\mathbf{p}_1\mk},\nonumber\\
&&\hat{T}^{R(A)}_{\mk\mathbf{p},LR}(\omega)\approx\hat{t}^{\dagger}_{\mk\mathbf{p}}+\sum_{\mk_1,\mathbf{p}_1}\hat{t}^{\dagger}_{\mk\mathbf{p}_1}\hat{g}^{R(A)}_{\mathbf{p}_1,RR}(\omega)\hat{t}_{\mathbf{p}_1\mk_1}\hat{g}^{R(A)}_{\mk_1,LL}(\omega)\hat{t}^{\dagger}_{\mk_1\mathbf{p}},\nonumber\\
&&\hat{G}^{R(A)}_{\mk\mathbf{p},LR}(\omega)\approx\hat{g}^{R(A)}_{\mk,LL}(\omega)\hat{t}^{\dagger}_{\mk\mathbf{p}}\hat{g}^{R(A)}_{\mathbf{p},RR}(\omega).\nonumber
\eeq
By substituting above into Eqs.~\eqref{eq:main1}-\eqref{eq:main3}, we obtain
\beq
&&I_N=\frac{\mu}{g}\Big[2t_{cc}\sin\Delta\phi -8\pi t^2_{nc}\Psi(0)\sin2\Delta\phi\nonumber\\
&&  +2(2\pi)^2t^2_{nc}t_{nn}\{\Phi^2(0)\sin\Delta\phi+\Psi^2(0)\sin3\Delta\phi \} \nonumber\\
&& -8(2\pi)^3t^2_{nc}t^2_{nn}\{2\Phi(0)\Psi^2(0)\sin2\Delta\phi+\Psi^3(0)\sin4\Delta\phi \}\Big]\nonumber\\
&&+O(t_{\mathbf{p}\mk}^5).
\label{eq:junction-current}
\eeq
Here, $\Phi$ and $\Psi$ are same as ones used in the point contact case (see also Appendix A).
We note that the first two terms of the right hand side in Eq.~\eqref{eq:junction-current} corresponds to the linear response theory 
and are consistent with~\cite{PhysRevA.64.033610}~\footnote{To be precise, our $\Delta\phi$ has the opposite sign to one used in~\cite{PhysRevA.64.033610}.}.
The terms proportional to $t^2_{nc}t_{nn}$ and $t^2_{nc}t^2_{nn}$ are the new results obtained owing to the nonlinear response analysis in this work.
Finally, we point out that  it is also possible to obtain a current expression
 including higher order in $t_{\mathbf{p}\mk}$ by incorporating higher tunneling processes in $\hat{T}_{LR(RL)}$ and $\hat{G}_{LR}$.

\section{Discussions}
By using the tunneling Hamiltonian and Bogoliubov theory, together with the Keldysh formalism, we have derived the DC Josephson current expression of  a weakly-interacting BEC.
We applied the derived formula to quantum point contact and planar junction systems and discussed how the current-phase relations 
are modified from the conventional sinusoidal Josephson relation as increasing the tunnel coupling.

The formulation presented in this work is based on the previous work~\cite{PhysRevResearch.2.023284},
where AC Josephson and direct currents in a BEC quantum point contact system  have been examined
under the presence of chemical potential and temperature biases.
As technical differences, this work uses Green's functions without the phonon approximation and
 takes into account the momentum dependence in the tunnel coupling, 
which allow to obtain more general current expressions including the planar junction.
In addition, a connection between AC and DC Josephson currents can be understood as follows.
In the presence of a chemical potential bias, oscillating current as a function of time that contain both sine and cosine components is 
generated. As shown in~\cite{PhysRevResearch.2.023284},  the amplitudes in the cosine components vanishes in the small bias limit
while those in the sine components remain nonzero, which is also consistent with the AC Josephson current in a fermionic superconductor.
Then, the zero bias limit of the AC Josephson current 
must be converged to the DC Josephson current by replacing the time oscillating argument with the relative phase.

Our formulation can also be applied to the so-called internal Josephson effect~\cite{PhysRevLett.105.204101},
where a two-component BEC with an internal coupling is concerned.
In such a system,  the coupling 
\beq
\int d^3x\Omega[\psi^{\dagger}_{\uparrow}\psi_{\downarrow}+\psi^{\dagger}_{\downarrow}\psi_{\uparrow} ],
\label{eq:rabi}
\eeq
with a Rabi frequency $\Omega$ plays a role of the tunneling term.
In the typical situation that the spatial dependence of  $\Omega$ is negligible, however, the above coupling term
only produce the momentum-conserving process between $\uparrow$ and $\downarrow$.
Thus, the dominant DC Josephson current in this situation can be obtained by $t_{nc}=t_{nn}=0$ in the planar junction case. 
Due to the absence of the momentum-nonconserving processes, the renormalization of the tunnel coupling does not occur.
Thus, it turns out that the internal Josephson current is described by the conventional sinusoidal curve, which is consistent with the Gross-Pitaevskii analysis~\cite{pitaevskii2016}.

At the same time, when $\Omega$ in Eq.~\eqref{eq:rabi} contains a nonnegligible spatial variation,
higher-order tunneling processes are generated.
Recently, such a coupling term is realized in  a mixture system of $^1$S$_0$  and $^3$P$_0$ with Yb atoms~\cite{ono2021}.
If the similar system can also be realized in bosonic mixtures,
one may harness Eqs.~\eqref{eq:main1}-\eqref{eq:main3} obtained in this paper.

We also point out that the tunneling Hamiltonian approach can also be applied to the heat current.
The heat current operator is defined as
\beq
I_Q=-\dot{H}_L=i[H_L,H_T].
\eeq
The commutator above consist of two parts: one between kinetic energy and tunneling terms and one between
interaction energy and tunneling term.
Especially, the latter part leads to a term quartic in the field operators whose analyses beyond the linear response regime is not so simple.
To obtain a more convenient expression, we consider the Heisenberg equation for  $b_{\mk,L}$
\beq
\dot{b}_{\mk,L}=-i[b_{\mk,L},H].
\eeq
The explicit evaluation of the above commutator shows that the heat current operator can be rewritten as~\cite{PhysRevLett.118.105303}
\beq
I_Q=-\sum_{\mathbf{p},\mk}( t_{\mathbf{p}\mk}b^{\dagger}_{\mathbf{p},R}\dot{b}_{\mk,L}+ t^*_{\mathbf{p}\mk}\dot{b}^{\dagger}_{\mk,L}b_{\mathbf{p},R} ).
\eeq
As in the case of the particle current, the average heat current is expressed in terms of lesser Green's function as follows:
\beq
I_Q=\sum_{\mathbf{p},\mk}\lim_{\tau'\to\tau}\text{Tr}\Big[\begin{pmatrix}
-t_{\mathbf{p}\mk}e^{-i\Delta\phi} & 0\\
0 & -t_{\mathbf{p}\mk}e^{i\Delta\phi}
\end{pmatrix}\nonumber\\
\ \ \times i\partial_{\tau} \hat{G}^<_{\mk\mathbf{p},LR}(\tau,\tau') \Big].
\eeq
Since the form of $\hat{G}^<$ has already been obtained in the analyses of the particle current in this paper,
it is straightforward to confirm that
the heat current originating from the coupling associated with the condensation elements is obtained as
\beq
I_Q=\int_{-\infty}^{\infty}\frac{d\omega}{2\pi}\omega[I^{\text{odd}}_{N}(\omega)+I_N^{\text{even}} (\omega)],
\eeq
where $I_N^{\text{odd}}$ and $I_N^{\text{even}}$ are given by Eqs.~\eqref{eq:main2} and~\eqref{eq:main3}.
Since $\hat{g}^<_{\mathbf{k}}(\omega)\propto \delta(\omega)$, we find
\beq
I_Q\approx0.
\eeq
This physically means that the tunneling processes associated with condensation elements do not carry heat.
Thus, if any, a nonzero heat current comes from the Bogoliubov-mode tunneling, 
which contains the Lee-Huang-Yang factor $\sqrt{va^3}$ with the $s$-wave scattering length $a$~\cite{PhysRevA.64.033610,PhysRevResearch.2.023284} and therefore
is expected to gives rise to a small effect in a weakly-interacting
BEC~\footnote{When it comes to the Bogoliubov-mode tunneling,
the momentum dependence in $t_{\mathbf{p}\mk}$ plays a crucial role in a temperature dependence in the current.
Indeed, the particle current contribution of the Bogoliubov-mode tunneling is proportional to $T^2$ in the quantum point contact 
but is proportional to $T^4$ in the planar junction. Technically, this difference originates from presence or absence of 
the momentum conservation along the interface.}.

So far, we have analyzed the two-terminal DC Josephson current where each reservoir is assumed to be uniform.
By using box potential techniques available in ultracold atomic gases~\cite{navon2021quantum},
the current in such a geometry can directly be measured and in fact has been measured in Ref.~\cite{luick2020}.
At the same time, by taking into account the fact that all experiments do not always implement such a scheme,
we briefly comment on trapped systems. Usually, spatial inhomogeneities in ultracold atomic gases can be explained with the local density approximation~\cite{inguscio2007}.
If the spatial inhomogeneity in each reservoir is indeed enough smooth that the local density approximation is available, 
the analyses presented in this paper can be applied to trapped systems.
In the case of the point contact system, the current is purely expressed in terms of local Green's functions 
at the center of the trap. Thus,  Eq.~\eqref{eq:leading} can directly be applied to the trapped case, provided
that $\mu$ or $v$ is now one at the center of the trap. Such an analysis has indeed been used in Ref.~\cite{husmann2015}.
In the case of the planar junction system, on the other hand, one may locally introduce the hopping amplitude $t_{\mathbf{p}\mk}(\mathbf{r})$.
As discussed in~\cite{PhysRevA.100.063601}, the total current should then be determined from the spatial integration of the corresponding local current.
 
Finally, we mention possible experiments to measure the DC Josephson current.
In ultracold atomic gases, such a current can be measured by making  current-biased junctions~\cite{barone1982},
which have been realized in a bosonic system~\cite{levy2007ac} and in a molecular BEC system~\cite{kwon2020,PhysRevLett.126.055301}.
There, an external current in current-biased junctions is generated by shifting a trapping potential~\cite{levy2007ac} or potential barrier~\cite{kwon2020,PhysRevLett.126.055301}.
It is then interesting to see the effects of higher order tunneling processes, which have already been done in a superconducting quantum point contact in condensed matter~\cite{bretheau2012}

\section*{Acknowledgments}
The author thanks J.-P. Brantut, T. Esslinger, P. Fabritius, T. Giamarchi, S. H\"{a}usler,
 M.-Z. Huang,  
J. Mohan, K. Nakata, Y. Nishida,  K. Ono, Y. Takahashi, M. Talebi, A.-M. Visuri, S. Wili, and S.-K. Yip for discussions. 
The author is supported by MEXT Leading Initiative for Excellent Young Researchers,
JSPS KAKENHI Grant No. JP21K03436, and Matsuo Foundation.

\appendix

\section{Uncoupled Green's functions}

Here, we derive formulae of uncoupled Green's functions used in the main text.

Due to the absence of the tunnel coupling, uncoupled Green's functions are diagonal in the momentum index 
$\hat{g}_{\mathbf{p}\mk}=\hat{g}_{\mathbf{p}}\delta_{\mathbf{p}\mk}$.
When it comes to retarded and advanced Green's functions, moreover, they are diagonal in the reservoir index
$\hat{g}^{R(A)}_{\alpha\beta}=\hat{g}^{R(A)}_{\alpha\alpha}\delta_{\alpha\beta}$, since   the commutator between the condensation elements $\mk=\mathbf{0}$ vanishes.
Thus, the retarded Green's functions in the frequency space within the Bogoliubov theory is given by
\begin{widetext}
\beq
\hat{g}^{R}_{\mk,\alpha\alpha}(\omega)&&=-i\int_{-\infty}^{\infty}d\tau e^{i\omega\tau}\theta(\tau)
\begin{pmatrix}
\langle[b_{\mathbf{k},\alpha}(\tau),b^{\dagger}_{\mathbf{k},\alpha}(0)] \rangle_0 & \langle[b_{\mathbf{k},\alpha}(\tau),b_{\mathbf{-k},\alpha}(0)] \rangle_0\\
\langle[b^{\dagger}_{\mathbf{-k},\alpha}(\tau),b^{\dagger}_{\mathbf{k},\alpha}(0)] \rangle_0 & \langle[b^{\dagger}_{\mathbf{k},\alpha}(\tau),b_{\mathbf{k},\alpha}(0)] \rangle_0
\end{pmatrix}\nonumber\\
&&=\begin{pmatrix}
\frac{u_k^2}{\omega-E_k+i0^+}-\frac{v_k^2}{\omega+E_k+i0^+} &\frac{u_kv_k}{\omega-E_k+i0^+}-\frac{u_kv_k}{\omega+E_k+i0^+} \\
\frac{u_kv_k}{\omega-E_k+i0^+}-\frac{u_kv_k}{\omega+E_k+i0^+} & \frac{v_k^2}{\omega-E_k+i0^+}-\frac{u_k^2}{\omega+E_k+i0^+}
\end{pmatrix},
\eeq
where $\langle\cdots\rangle_0$ means an average without $H_T$.
We note that since the same chemical potential between reservoirs is assumed, $\hat{g}^R$ does not depend on $\alpha$.
Similarly, advanced Green's function can simply be obtained by using the relation $\hat{g}^A_{\mk,\alpha\alpha}(\omega)=[\hat{g}^R_{\mk,\alpha\alpha}(\omega)]^{\dagger}$.
In contrast, lesser Green's function contain an off-diagonal  component in the reservoir index, which is associated with the presence of a BEC. 
By using the Bogoliubov theory, lesser Green's function is determined as
\beq
&&\hat{g}^<_{\mk,\alpha\beta}(\omega)=-i\int_{-\infty}^{\infty}d\tau e^{i\omega\tau}
\begin{pmatrix}
\langle b^{\dagger}_{\mathbf{k},\alpha}(0)b_{\mathbf{k},\alpha}(\tau) \rangle_0 & \langle b_{\mathbf{-k},\alpha}(0)b_{\mathbf{k},\alpha}(\tau) \rangle_0\\
\langle b^{\dagger}_{\mathbf{k},\alpha}(0)b^{\dagger}_{\mathbf{-k},\alpha}(\tau) \rangle_0 & \langle b_{\mathbf{k},\alpha}(0)b^{\dagger}_{\mathbf{k},\alpha}(\tau) \rangle_0
\end{pmatrix}\nonumber\\
&&=\begin{cases}
-\frac{2\pi\mu i}{g}\delta(\omega)\begin{pmatrix} 1 & 1\\ 1 & 1
\end{pmatrix} \ \ \ (\mk=\mathbf{0})\\
\delta_{\alpha\beta}\begin{pmatrix}
-2\pi i n_{\alpha}(\omega)[u_k^2\delta(\omega-E_k)-v_k^2\delta(\omega+E_k)] & 2\pi i n_{\alpha}(\omega)u_kv_k[\delta(\omega-E_k)-\delta(\omega+E_k)]  \\
2\pi i n_{\alpha}(\omega)u_kv_k[\delta(\omega-E_k)-\delta(\omega+E_k)] & 2\pi i n_{\alpha}(\omega)[u_k^2\delta(\omega+E_k)-v_k^2\delta(\omega-E_k)] 
\end{pmatrix} \ \ (\mk\ne\mathbf{0})
\end{cases}
\label{eq:lesser}
\eeq
with the Bose distribution function
$n_{\alpha}(\omega)=\frac{1}{e^{\omega/T_{\alpha}}-1}.$
We note that $\hat{g}^<_{\mathbf{0},\alpha\beta}$ takes the same expression irrespective of reservoir indices $\alpha,\beta$, and
the factor $\delta(\omega)$ appearing in $\mk=\mathbf{0}$ component represent the fact that the condensation elements do not evolve in time~\cite{fetter2012}.
 
\end{widetext}
\subsection{Quantum point contact case}
In the case of quantum point contact transport, uncoupled Green's functions that take the sum over momentum index are concerned.
We note that they are essentially local Green's functions at origin since $\langle T_c\psi(\mathbf{0},\tau) \psi^{\dagger}(\mathbf{0},\tau')\rangle=\sum_{\mk}\langle T_c b_{\mk}(\tau) b_{\mk}^{\dagger}(\tau')\rangle$
with $\psi(\mathbf{0})=\sum_{\mk}b_{\mk}$.
Since local Green's function may contain  a UV divergence, we introduce a momentum cutoff $\Lambda$ to regularize the problem.
The introduction of the momentum cutoff is physically sound in that after all the tunneling Hamiltonian describes low-energy transport of the system.
Then, local regarded Green function at the frequency space is given by
\beq
\hat{g}^R_{\alpha\alpha}(\omega)&&=\sum_{\mathbf{k}}^{\Lambda}\hat{g}^R_{\mk,\alpha\alpha}(\omega)\nonumber\\
&&\equiv\begin{pmatrix}
g^R_1(\omega) & f^R(\omega)\\
f^R(\omega) & g^R_2(\omega)
\end{pmatrix}.
\label{eq:local-ret}
\eeq
where
\beq
g^R_1(\omega)&&=\Phi(\omega)-i\frac{\rho(\omega)}{2},\\
g^R_2(\omega)&&=\Phi(-\omega)+i\frac{\rho(-\omega)}{2},\\
f^R(\omega)&&=\Psi(\omega)-i\frac{\sigma(\omega)}{2}.
\eeq
Here, $\Phi$ and $\Psi$ are real parts of retarded Green's function that are given by
\beq
&&\Phi(\omega)=-\frac{\sqrt{2}\mu^2 }{c^3\pi^2}
\Big[\frac{c\Lambda}{\sqrt{2}\mu }\nonumber\\
&&+\frac{\pi(\bar{\omega}-1 )}{4\sqrt{1+\bar{\omega}^2}}\Big\{\sqrt{1+\sqrt{1+\bar{\omega}^2}}-
\sqrt{1-\sqrt{1+\bar{\omega}^2}}\Big\}\nonumber\\
&&-\frac{\pi\bar{\omega}^2}{4\sqrt{1+\bar{\omega}^2}}\Big\{\frac{1}{\sqrt{1+\sqrt{1+\bar{\omega}^2}}}-
\frac{1}{\sqrt{1-\sqrt{1+\bar{\omega}^2}}}
\Big\}\Big],\nonumber\\
\eeq
\beq
\Psi(\omega)&&=\frac{\sqrt{2}\mu^2}{c^3\pi^2}\Big[ \frac{\pi}{4\sqrt{1+\bar{\omega}^2}}\Big\{\sqrt{1+\sqrt{1+\bar{\omega}^2}}\nonumber\\
&&-\sqrt{1-\sqrt{1+\bar{\omega}^2}}\Big\} \Big],
\eeq
with  $\bar{\omega}=\omega/\mu$.
In addition, $\rho$ and $\sigma$ are the density of states associated with the imaginary parts of retarded Green's function. 
For $\omega>0$, they are determined as
\beq
\rho(\omega)=\frac{ k_0(\epsilon_{k_0}+\mu+E_{k_0})}{2\pi c^2(1+\epsilon_{k_0}/\mu)}\theta(\Lambda-k_0),
\eeq
\beq
\rho(-\omega)=-\frac{ k_0(\epsilon_{k_0}+\mu-E_{k_0})}{2\pi c^2(1+\epsilon_{k_0}/\mu)}\theta(\Lambda-k_0),
\eeq
\beq
\sigma(\omega)=-\frac{\mu k_0}{2\pi c^2(1+\epsilon_{k_0}/\mu)}\theta(\Lambda-k_0),
\eeq
\beq
\sigma(-\omega)=\frac{\mu k_0}{2\pi c^2(1+\epsilon_{k_0}/\mu)}\theta(\Lambda-k_0),
\eeq
where
\beq
k_0=\sqrt{2M\sqrt{\mu^2+\omega^2}-2M\mu}.
\eeq
Again, advanced Green function can be determined from
$\hat{g}^A(\omega)=[\hat{g}^{R}(\omega)]^{\dagger}$.
Finally, uncoupled lesser Green function in frequency space is obtained as
\beq
&&\hat{g}^<_{\alpha\beta}(\omega)=\sum_{\mk}\hat{g}^<_{\mk,\alpha\beta}(\omega)
\nonumber\\
&&=-\frac{2\pi i\mu}{g}\delta(\omega)
\begin{pmatrix}
1 &1 \\
1 & 1
\end{pmatrix}-i\delta_{\alpha,\beta}n_{\alpha}(\omega)
\begin{pmatrix}
\rho(\omega)&\sigma(\omega) \\
\sigma(\omega) &-\rho(-\omega)
\end{pmatrix}.\nonumber\\
\eeq

\subsection{Planar junction case}
For calculation of the dominant DC Josephson current in the planar junction,
it is important to evaluate
the following form of retarded or advanced Green's function:
\beq
\hat{\bar{g}}^{R(A)}_{\alpha\alpha}(\omega)=\sum_{k}k^2\hat{g}^{R(A)}_{k,\alpha\alpha}(\omega),
\eeq
where  $k$ is the momentum perpendicular to the interface and $\mathbf{k}_{\parallel}=\mathbf{0}$ is assumed above.
Since $\hat{g}^{R(A)}_{-k,\alpha\alpha}(\omega)=\hat{g}^{R(A)}_{k,\alpha\alpha}(\omega)$, we see that the similar calculation to one in the point contact case is allowed.
A straightforward calculation leads
\beq
\hat{\bar{g}}^R_{\alpha\alpha}(\omega)=2\pi\begin{pmatrix}
g^R_1(\omega) & f^R(\omega)\\
f^R(\omega) & g^R_2(\omega)
\end{pmatrix},
\eeq
where $g^R_1$, $g^R_2$, and $f^R$ are introduced in Eq.~\eqref{eq:local-ret}.
Thus, retarded and advanced Green's functions relevant to the dominant DC Josephson effect in the planar junction turns out to be obtained by multiplying ones in the point contact by $2\pi$.

\section{DC Josephson current in an $s$-wave superconductor}
In this appendix, we show a derivation of  the DC Josephson current in conventional BCS superconductors by means of the tunneling Hamiltonian,
which has already be done by using the fluctuation-dissipation relation in the total system~\cite{PhysRevB.51.3743}.
For future reference, here we give another derivation without use of the fluctuation-dissipation relation in the total system,
which allows us to extend the system with a temperature imbalance between reservoirs.

In what follows, we assume that the tunneling amplitude is momentum-independent, which is usually done in fermionic systems~\cite{mahan2013}.
Then, the analysis is reduced to one used in the point contact system.
Namely, we can consider the following tunneling term
\beq
H_T=\hat{\psi}^{\dagger}_{R}(\mathbf{0})\hat{t}\hat{\psi}_{L}(\mathbf{0})+\hat{\psi}^{\dagger}_{L}(\mathbf{0})\hat{t}^{\dagger}\hat{\psi}_{R}(\mathbf{0}),
\eeq
where we introduce the Nambu spinor
\beq
\hat{\psi}_{\alpha}=\begin{pmatrix}
\psi_{\uparrow,\alpha}\\
\psi_{\downarrow,\alpha}
\end{pmatrix},
\eeq
and the tunneling matrix
\beq
\hat{t}=\begin{pmatrix}
-te^{-i\Delta\phi/2} & 0\\
0 & te^{i\Delta\phi/2}
\end{pmatrix},
\eeq
with the relative phase $\Delta\phi$.
We note that $\psi_{\uparrow/\downarrow}$ obeys the fermionic anti-commutation relation.
The presence of $e^{\pm i\Delta\phi/2}$ in the above tunneling matrix comes from the fact that
 Cooper pairs carry the phase $\phi_{L(R)}$ and fermions carry half of it.
By introducing local lesser Green's function,
\beq
&&\hat{G}^<_{LR}(\tau,\tau')\nonumber\\
&&=i
\begin{pmatrix}
 \langle \psi^{\dagger}_{\uparrow,R}(\mathbf{0},\tau') \psi_{\uparrow,L}(\mathbf{0},\tau)\rangle & \langle \psi_{\downarrow,R}(\mathbf{0},\tau') \psi_{\uparrow,L}(\mathbf{0},\tau)\rangle\\
\langle \psi^{\dagger}_{\uparrow,R}(\mathbf{0},\tau') \psi^{\dagger}_{\downarrow,L}(\mathbf{0},\tau)\rangle & \langle \psi_{\downarrow,R}(\mathbf{0},\tau') \psi^{\dagger}_{\downarrow,L}(\mathbf{0},\tau)\rangle
\end{pmatrix},
\nonumber\\
\eeq
the particle current is expressed as
\beq
I_N=-\text{Tr}\Big[\hat{\sigma}_z\Big(\hat{t}\hat{G}^<_{LR}(\tau,\tau)-\hat{t}^{\dagger}\hat{G}^<_{RL}(\tau,\tau)\Big)\Big].
\eeq

We now evaluate $I_N$ within the BCS theory.
Then, uncoupled retarded (advanced) Green's function in the frequency-space representation is determined  as~\cite{PhysRevB.51.3743}
\beq
\hat{g}^R_{\alpha\alpha}(\omega)&&=\frac{1}{W\sqrt{\Delta^2_{\alpha}-(\omega+i0^+)^2}}
\begin{pmatrix}
-(\omega+i0^+) &\Delta_{\alpha} \\  
\Delta_{\alpha} & -(\omega+i0^+)
\end{pmatrix}\nonumber\\
&&=\begin{pmatrix}
g^R_{1,\alpha}(\omega) & g^R_{2,\alpha}(\omega) \\  
g^R_{2,\alpha}(\omega) & g^R_{1,\alpha}(\omega)
\end{pmatrix},
\eeq
where $\hat{g}^A_{\alpha\alpha}(\omega)=[\hat{g}^{R}_{\alpha\alpha}(\omega)]^{\dagger}$,  $W=1/[\pi\rho(\epsilon_F)]$ with the density of state at the Fermi energy $\rho(\epsilon_F)$,
and $\Delta_{\alpha}$ is the superconducting energy gap in reservoir $\alpha$.
We note that in order to obtain the expression above, we use the approximation that the density of states in free fermions is replaced by a constant value determined at the Fermi energy.
Similarly, uncoupled lesser Green's function in the frequency space is obtained from the fluctuation-dissipation relation in each reservoir,
\beq
\hat{g}^<_{\alpha\alpha}(\omega)=-n_{F,\alpha}(\omega)[\hat{g}^R_{\alpha\alpha}(\omega)-\hat{g}^A_{\alpha\alpha}(\omega)],
\label{eq:FDR}
\eeq
with the Fermi distribution function $n_{F,\alpha}(\omega)=\frac{1}{e^{\omega/T_{\alpha}}+1 }$.
We note that the above usage of the fluctuation-dissipation relation is limited within each reservoir
and we do not assume that the total system is in equilibrium.

Based on uncoupled Green's functions introduced above, we can obtain Dyson equations relevant to the Josephson current.
Dyson equations for retarded and advanced components are given by
 \beq
\hat{G}^{R(A)}_{LR}
&&=\hat{G}^{R(A)}_{LL}\hat{t}^{\dagger}\hat{g}^{R(A)}_{RR},\\
\hat{G}^{R(A)}_{RL}
&&=\hat{G}^{R(A)}_{RR}\hat{t}\hat{g}^{R(A)}_{LL},\\
\hat{G}^{R(A)}_{LL}
&&=\hat{g}^{R(A)}_{LL}+\hat{G}^{R(A)}_{LR}\hat{t}\hat{g}^{R(A)}_{LL},\\
\hat{G}^{R(A)}_{RR}
&&=\hat{g}^{R(A)}_{RR}+\hat{G}^{R(A)}_{RL}\hat{t}^{\dagger}\hat{g}^{R(A)}_{RR},
\eeq
On the other hand, Dyson equations for the lesser component are
\beq
\hat{G}^<_{RL}&&=(\hat{1}+\hat{G}^R_{RL}\hat{t}^{\dagger})\hat{g}^<_{RR}\hat{t}\hat{G}^A_{LL}
+\hat{G}^R_{RR}\hat{t}\hat{g}^<_{LL}(\hat{1}+\hat{t}^{\dagger}\hat{G}^A_{RL}),\nonumber\\ \\
\hat{G}^<_{LR}&&=(\hat{1}+\hat{G}^R_{LR}\hat{t})\hat{g}^<_{LL}\hat{t}^{\dagger}\hat{G}^A_{RR}
+\hat{G}^R_{LL}\hat{t}^{\dagger}\hat{g}^<_{RR}(\hat{1}+\hat{t}\hat{G}^A_{LR}).\nonumber\\
\eeq
By using Eq.~\eqref{eq:FDR}, 
they are rewritten as
\begin{widetext}
\beq
\hat{G}^<_{RL}&&=-n_{F,R}(\hat{1}+\hat{G}^R_{RL}\hat{t}^{\dagger})(\hat{g}^R_{RR}-\hat{g}^A_{RR})\hat{t}\hat{G}^A_{LL}
-n_{F,L}\hat{G}^R_{RR}\hat{t}(\hat{g}^R_{LL}-\hat{g}^A_{LL})(\hat{1}+\hat{t}^{\dagger}\hat{G}^A_{RL}),\nonumber\\
&&=-n_{F,R}\Big[\hat{G}^R_{RR}\hat{t}\hat{G}^A_{LL}-(\hat{1}+\hat{G}^R_{RL}\hat{t}^{\dagger})\hat{G}^A_{RL}\Big]
-n_{F,L}\Big[\hat{G}^R_{RL}(\hat{1}+\hat{t}^{\dagger}\hat{G}^A_{RL}) -\hat{G}^R_{RR}\hat{t}\hat{G}^A_{LL} \Big], \nonumber\\
\eeq
\beq
\hat{G}^<_{LR}&&=-n_{F,L}(\hat{1}+\hat{G}^R_{LR}\hat{t})(\hat{g}^R_{LL}-\hat{g}^A_{LL})\hat{t}^{\dagger}\hat{G}^A_{RR}
-n_{F,R}\hat{G}^R_{LL}\hat{t}^{\dagger}(\hat{g}^R_{RR}-\hat{g}^A_{RR})(\hat{1}+\hat{t}\hat{G}^A_{LR}),\nonumber\\
&&=-n_{F,L}\Big[\hat{G}^R_{LL}\hat{t}^{\dagger}\hat{G}^A_{RR}-(\hat{1}+\hat{G}^R_{LR}\hat{t})\hat{G}^A_{LR}\Big]
-n_{F,R}\Big[\hat{G}^R_{LR}(\hat{1}+\hat{t}\hat{G}^A_{LR}) -\hat{G}^R_{LL}\hat{t}^{\dagger}\hat{G}^A_{RR} \Big]. \nonumber\\
\eeq
\end{widetext}
The above expressions may also be useful to discuss the Josephson current in the presence of a temperature bias.
In what follows, we focus on the situation that two reservoirs have an equal temperature, that is, $n_{F,L}(\omega)=n_{F,R}(\omega)\equiv n_F(\omega)$.
Then, the above expressions are further simplified and we obtain
\beq
\hat{G}^<_{RL}&&=-n_F\Big[\hat{G}^R_{RL}-\hat{G}^A_{RL}\Big],\\
\hat{G}^<_{LR}&&=-n_F\Big[\hat{G}^R_{LR}-\hat{G}^A_{LR}\Big].
\eeq
Therefore, the current is expressed as
\beq
I_N&&=\int_{-\infty}^{\infty} \frac{d\omega}{2\pi}\text{Tr}\Big[\hat{\sigma}_z\hat{t}\{\hat{G}^R_{LR}-\hat{G}^A_{LR} \} -\hat{\sigma}_z\hat{t}^{\dagger}\{\hat{G}^R_{RL}-\hat{G}^A_{RL} \} \Big]n_F, \nonumber\\
\eeq
which is identical to one obtained from the fluctuation-dissipation relation in the total system~\cite{PhysRevB.51.3743},
since the same temperature between reservoirs is assumed.

To obtain a simpler form of the current, we now evaluate $\hat{G}^{R(A)}_{RL}$ and $G^{R(A)}_{LR}$ under $\Delta_L=\Delta_R\equiv \Delta$.
From the Dyson equations for retarded and advanced components, we obtain
\beq
\hat{G}^{R(A)}_{LR}(\omega)=\hat{g}^{R(A)}_{LL}(\omega)\hat{t}^{\dagger}\hat{g}^{R(A)}_{RR}(\omega)\hat{D}^{R(A)}_{LR}(\omega),\\
\hat{G}^{R(A)}_{RL}(\omega)=\hat{g}^{R(A)}_{RR}(\omega)\hat{t}\hat{g}^{R(A)}_{LL}(\omega)\hat{D}^{R(A)}_{RL}(\omega),
\eeq
where
\beq
\hat{D}^{R(A)}_{LR}(\omega)=[\hat{1}-\hat{t}\hat{g}^{R(A)}_{LL}(\omega)\hat{t}^{\dagger}\hat{g}^{R(A)}_{RR}(\omega)]^{-1},\\
\hat{D}^{R(A)}_{RL}(\omega)=[\hat{1}-\hat{t}^{\dagger}\hat{g}^{R(A)}_{RR}(\omega)\hat{t}\hat{g}^{R(A)}_{LL}(\omega)]^{-1}.
\eeq
\begin{widetext}
From $2\times2$ matrix inversion,
$\hat{D}^{R(A)}_{LR}(\omega)$ and $\hat{D}^{R(A)}_{RL}(\omega)$ are solved as
\beq
\hat{D}^{R(A)}_{LR}(\omega)=\frac{1}{D^{R(A)}(\omega)}
\begin{pmatrix}
 1-t^2(g^{R(A)}_1)^2+t^2(g^{R(A)}_2)^2e^{i\Delta\phi}  & -g^{R(A)}_1g^{R(A)}_2t^2[1-e^{-i\Delta\phi}]  \\
 -g^{R(A)}_1g^{R(A)}_2t^2[1-e^{i\Delta\phi}]  & 1-t^2(g^{R(A)}_1)^2+t^2(g^{R(A)}_2)^2e^{-i\Delta\phi}
\end{pmatrix},
\eeq
\beq
\hat{D}^{R(A)}_{RL}(\omega)=\frac{1}{D^{R(A)}(\omega)}
\begin{pmatrix}
 1-t^2(g^{R(A)}_1)^2+t^2(g^{R(A)}_2)^2e^{-i\Delta\phi}  & -g^{R(A)}_1g^{R(A)}_2t^2[1-e^{i\Delta\phi}]  \\
 -g^{R(A)}_1g^{R(A)}_2t^2[1-e^{-i\Delta\phi}]  & 1-t^2(g^{R(A)}_1)^2+t^2(g^{R(A)}_2)^2e^{i\Delta\phi}
\end{pmatrix},
\eeq
where
\beq
D^{R(A)}(\omega)&&=-\frac{(1+2t^2/W^2+t^4/W^4 )}{\Delta^2-(\omega\pm i0^+)^2}[(\omega\pm i0^++\omega_{ABS})
(\omega\pm i0^+-\omega_{ABS})].
\eeq
Here we introduce the Andreev bound state frequency:
\beq
\omega_{ABS}=\Delta\sqrt{1-{\cal T}\sin^2(\Delta\phi/2)},
\eeq
with the transmission parameter
\beq
{\cal T}=\frac{4t^2/W^2 }{(1+t^2/W^2)^2}.
\eeq
Notice that $0\le {\cal T}\le1$.
Now that $\hat{D}^{R(A)}_{LR}$ and $\hat{D}^{R(A)}_{RL}$ are determined, we can obtain $\hat{G}^{R(A)}_{LR}$ and $\hat{G}^{R(A)}_{RL}$.
A straightforward calculation reveals that the DC Josephson current is given by
\beq
I_N&&=\frac{2it^2\sin\Delta\phi}{\pi}\int_{-\infty}^{\infty}d\omega n_F(\omega)\Big[\frac{g_2^R(\omega)g_2^R(\omega)}{D^R(\omega)}
-\frac{g_2^A(\omega)g_2^A(\omega)}{D^A(\omega)} \Big]\nonumber\\
&&=-\frac{{\cal T}\Delta^2i\sin\Delta\phi}{4\pi\omega_{ABS}}\int_{-\infty}^{\infty}d\omega n_F(\omega)\Big[\frac{1}{\omega-\omega_{ABS}+i0^+}-\frac{1}{\omega+\omega_{ABS}+i0^+} 
-\frac{1}{\omega-\omega_{ABS}-i0^+}+\frac{1}{\omega+\omega_{ABS}-i0^+}\Big]\nonumber\\
&&=\frac{{\cal T}\Delta^2\sin\Delta\phi}{2\omega_{ABS}}\tanh\frac{\omega_{ABS}}{2T}.
\eeq
The final expression above is the celebrated DC Josephson current in  an $s$-wave superconductor.
It is clear from the derivation above that the current is carried by the Andreev bound state.
We note that the essentially same expression can also be obtained by using the Bogoliubov-de-Gennes~\cite{furusaki,beenakker,PhysRevB.45.10563} and quasi-classical Green's function methods~\cite{zaitsev,PhysRevB.68.024511}.

\end{widetext}

%\bibliographystyle{apsrev4-2}
%\bibliography{reference}

%

\end{document}